East Asians with Internet Addiction: Prevalence Rates and Support Use Patterns

Stephen Wong and Cass Dykeman[1]
Oregon State University
104 Furman Hall
Corvallis, OR 97331-3502

*A Preprint*

Abstract

The issue of Internet addiction has become a serious social and health issue in East Asian countries. There are only a few treatment programs for Internet addiction, and their effectiveness with people from East Asian remains unclear. As support and treatment develop, it is necessary to understand cultural preferences for dealing with this concern. Using data from the East Asian Social Survey (EASS), this study examined preferred sources of assistance for help with internet use problems in four countries – China, Japan, South Korea, and Taiwan. Preferences for kin versus non-kin support, use of alternative medicine, and professional mental health assistance were examined, as were between-country differences in support preferences. The results indicate a strong preference for seeking assistance from close relatives, followed by non-kin support (i.e., close friends and co-participants in religious institutions), alternative medicine, and professional mental health services, respectively. While there is a strong preference for family support, over 80% of survey respondents were open to seeking formal or informal mental health support outside the family. There were some significant differences between countries, with South Koreans being more likely to seek non-kin support and professional support for internet addiction concerns compared to Chinese. These differences are discussed in the context of cultural and policy developments in East Asian countries. Findings suggest the need for a more holistic approach to treating low mental health concerns.

[1] Correspondence concerning this article should be addressed to Cass Dykeman at dykemanc@oregonstate.edu

# East Asians with Internet Addiction: Prevalence Rates and Support Use Patterns

As the fastest-growing addiction, Internet addiction has become a serious global issue, with some of the highest prevalence in East Asia (Ko, Yen, Chen, Chen, Wu, & Yen, 2006; Ko, Yen, Yen, Lin, & Yang, 2007; Mak et al., 2014; Sung, Lee, Noh, Park, & Ahn, 2013; Wu & Zhu, 2004). With such a widespread phenomenon, preferred treatment modality becomes an essential consideration for clinicians and researchers. The purpose of the present study is to examine where East Asians seek help for Internet addiction.

It is suggested that Internet addiction is associated with psychological and behavioral changes. One study has shown that Internet addiction can cause depression, anxiety, hostility, interpersonal sensitivity, and psychoticism (Dong, Lu, Zhou, & Zhao, 2011). The high prevalence of Internet addiction in East Asian countries has become a serious social and health issue (Ko et sl., 2006; Sung, Lee, Noh, Park, & Ahn, 2013). Unfortunately, there are only a few treatment programs for Internet addiction, and their effectiveness in the context of East Asian cultures remain unclear. In response to the increasing prevalence of Internet addiction and its adverse effects, it is necessary to understand the different forms of support in order to develop effective treatment modalities in the context of East Asian cultures.

When reviewing the literature on Internet-addiction issues and the usage of support resources among East Asians, five topics stand out for consideration: the definition of Internet addiction, the prevalence of Internet addiction in East Asian countries, professional Internet-addiction support, alternative assistance for persons with addictions, and social support. After these issues are examined, the research question that guided this study are presented.

Internet addiction is described as uncontrolled and dysfunctional use of the internet, which recent psychiatric literature recognized as one of the impulse-control disorders (Dell'Osso, Altamura, Allen, Marazziti, & Hollander, 2006; Shapira, Goldsmith, Keck, Khosla, & McElroy, 2000).

According to the American Psychiatric Association, Internet addiction is a compulsive-impulsive spectrum disorder that involves online and/or offline computer usage (Dell'Osso et al., 2006).  There are at least three subtypes: excessive gaming, sexual preoccupations, and e-mail/text messaging (Block, 2007).  All three subtypes share four components: (a) excessive use, which associates with a loss of sense of time or a neglect of basic drives; (b) withdrawal, which includes feelings of anger, tension, and/or depression when Internet usage is not accessible; (c) tolerance, such as the urge to pursue better equipment, software, or more hours of usage; and (d) negative repercussions, which include arguments, lying, social isolation, and fatigue (Beard & Wolf, 2001; Block, 2007).  This well-formulated definition helps researchers identify patients whose behavior suggests addiction.

With the increasing accessibility of the Internet, the number of Internet-addiction cases has also significantly increased.  Internet addiction has become a growing issue in East Asia.  For instance, Wu and Zhu (2004) identified 10.6% of Chinese college students as having an Internet addiction.  About 3% of Korean adolescents could be classified as high risk for Internet addiction (Sung et al., 2013).  Among this high-risk group, 66.7% of the adolescents reported feeling stressed and unhappy.  Another study with Taiwanese students revealed that the prevalence rate of Internet addiction in adolescents is 17.7% (Ko et al., 2007); it suggested that Taiwanese adolescents with Internet addiction were more likely to have substance-use experiences including tobacco, alcohol, or illicit drug use (Ko et al., 2006).  In Japan, the prevalence of problematic or addictive Internet use among students was as high as 48% (Mak et al., 2014).  Such a high prevalence of Internet addiction is a growing concern for the East Asian countries.  The development of professional may be needed to meet the needs of patients.

The availability of cultural-sensitive intervention programs for Internet addiction varies among East Asian countries.  There are several existing intervention models in China, where researchers have developed trials to apply psychoanalytic group intervention (Yang, Li, He, & Zhao, 2008), family therapy (Gong, Wang, Ye, & Liang, 2010), sports-exercise prescriptions (Zhang, 2009), and Naikan therapy (Su, Fang, Miller, & Wang, 2011).  The availability and accessibility of these interventions are limited, and they can be costly (Su, Fang, Miller, & Wang, 2011).  In addition to the traditional interventions, boot-camp-style rehab programs have emerged in both China and Korea (Koo, Wati, Lee, & Oh, 2011). The South Korean government has established about 140 counseling centers and developed treatment programs at almost 100 hospitals (Kim, 2008). In Japan, the government has developed "fasting camps" to help individuals with Internet addictions (Majumdar, 2013). An important limitation of the previous research on Internet-addiction interventions is its heavy focus on the pathological aspect of Internet addiction. Studies on professional interventions and their effectiveness are scarce. Additional resources are necessary in order to meet the ever-growing needs.

While professional assistance might be difficult to locate in most East Asian countries, some researchers and practitioners have been exploring other alternatives. For example, there is evidence showing that alternative treatments, such as electroacupuncture, may be effective in treating Internet addiction. One study suggests that electroacupuncture combined with psychologic interference can significantly improve the anxiety states of individuals with Internet addiction (Zhu, Jin, Zhong, Chen, & Li 2008). Another study shows that electroacupuncture in combination with psychointervention may improve the cognitive function of Internet-addicted patients (Zhu et al., 2012). While these studies seem to be promising for treating Internet addiction, they are in their infancy stages, and more research is necessary for development and implementation.

Other than professional support and alternative medicine, social support is one important factor in the recovery process for Internet-addicted individuals. Social support has a strong negative correlation with Internet addiction. Studies show that factors such as parenting attitudes, family communication, and family function are strongly associated with Internet addiction among South Korean adolescents (Cho, 2001; Kim, 2001; Nam, 2002; Park, Kim, & Cho, 2008). Similar studies in Taiwan and Hong Kong echoed the findings, showing a strong association between Internet addiction and parent-adolescent conflict and family function (Yen, Yen, Chen, Chen, & Ko, 2007; Yu & Shek, 2013). A study conducted with Japanese and Chinese college students indicated the association between perception of parents being less caring and more controlling with Internet addiction (Yang, Yamawaki, & Miyata, 2013). The availability of social support also has an important effect on the treatment of Internet addiction; however, we know little about the prevalence of such support sources among East Asians

(Lam. Peng, Mai, & Jing, 2009). Moreover, despite cultural similarities among Asians, research suggested that there are significant differences among Asian subgroups in attitudes toward mental health issues (Fung & Wong, 2007). One study found that Korean immigrants were less likely to endorse marital violence than their Chinese counterparts (Yoshioka, DiNoia, & Ullah, 2001). Another study suggested that Koreans are more likely to endorse "traditional, non-Western beliefs, including both 'non-Western physiological' and supernatural beliefs'" compared to Chinese (Fung & Wong, 2007). There is little quantitative analysis of how East Asian countries differ regarding the rate of support sources. The present study addresses this critical gap in the literature.

The present study was guided by eight research questions about persons reporting Internet addiction: 1) What is the usage rate of kin as a support resource? 2) Are there statistically significant differences in the use of kin as a support resource by country? 3) What is the usage rate of nonkin as a support resource? 4) Are there statistically significant differences in the use of non-kin as a support resources by country? 5) What is the usage rate of alternative medicine as a support resource? 6) Are there statistically significant differences in the use of alternative medicine as support resources by country? 7) What is the usage rate of professional mental health services as a support resource?, and 8) Are there statistically significant differences in the use of professional mental health services as support resources by country?

## Methods

This study employed a retrospective, cross-sectional observational analysis to determine between-country differences in the utilization of different types of health assistance and the prevalence of East Asians reporting Internet addiction. Data were derived from the East Asian Social Survey (EASS). The EASS is a biennial social survey project that serves as a cross-national network of the following four General Social Survey-type surveys in East Asia: (1) Chinese General Social Survey (CGSS); (2) Japanese General Social Survey (JGSS); (3) Korean General Social Survey (KGSS); and (4) Taiwan Social Change Survey (TSCS). The purpose of these surveys is to compare diverse aspects of social life in these regions. Survey information in this module focused on issues that affected overall health, such as specific conditions, physical functioning, aid received from family members or friends when needed, and lifestyle choices. While EASS assessed a number of variables, this study focused on internet addiction and various sources of assistance. The variables in this study were (a) presence of an Internet addiction, (b) use of kin as support for an Internet addiction, (c) use of non-kin as support for an internet addiction, (d) use of alternative medicines as support for an internet addiction, (e) use of professional mental health as support for an internet addiction, and (f) country. EASS utilized multistage sampling procedures to gather data: China, three-stage PPS; Japan, two-stage stratified random sampling (stratified by regional block and population size); South Korea, multistage area probability sampling; and Taiwan, three-stage stratified PPS sampling (PSU [township], village, and individual person).

**Participants.** The initial sample size of the 2010 Chinese General Social Survey was 5,370, with 3,866 responding; the sample was composed of Chinese aged 18 and above. The initial sample size of the 2010 Japanese General Social Surveys was 4,500, with 2,496 responding; the sample was composed of men and women aged 20–89 living in Japan. The initial sample size of the 2010 Korean General Social Survey was 2,500, with 1,576 responding; the sample was composed of adult citizens aged 18 and over who lived in households in South Korea. The initial sample size of the 2011 Taiwan Social Change Survey was 4,424, with 2,199 responding; the sample was composed of population registers. Of the 10,137 participants in this study, 351 reported having addiction issues; none of the 2,199 participants from Taiwan provided information on mental health.

A total of 351 respondents who reported internet addiction were selected in which consisted of a total 251 males and 100 females. Among the samples, 46.6% was between the ages of 20 to 29 and 27.7% was between the ages of 30 to 39. There was 54.9% of the sample reported never married and about 43.3% was reported married. The range of the number of years of education was 21 years with a mean of 13.29 years.

**Measures**. All data were drawn from items contained in the East Asian Social Survey (EASS; Iwai et al., 2010). The item numbers that appear in this subsection refer to the item numbers in the data codebook and files of EASS.

***Internet addiction.*** Internet addiction was assessed using a single item (V78 EASS 2010) asking for a respondent's self-report of others' perceptions of that the respondent's use was excessive: "Have you ever done or has anyone told you that you have done the following behavior excessively? Video/Internet games (including cell phone games)." The possible responses were "Yes (1)," recorded as "1," and "No (2)," recorded as "0." The validity and reliability of single-item measures and measures that assess third-party perceptions have been supported (Sanders & Williams, 2016). Sanders and Williams (2016) report that third-party perceptions of addictions loaded on the second factor component of the significant problematic internet/gaming use measure.

***Use of kin as a support resource.*** The use of kin as a support resource was assessed using a single item (V49 EASS 2010): "During the past 12 months, did your kin (family or relatives) do the following things for you when you needed it? If yes, how often?" The possible responses to this question were "Very often (1)," "Often (2)," "Sometimes (3)," "Seldom (4)," "Not at all (5)," "No, do not have such needs (6)," "No such persons available (7)," "Not asked (77)," and "Don't know/Refused (88)." Because the question specifically asked if the respondents received support when they needed it, response number 6 is not a useful indicator of frequency of support. Similarly, because the question is whether kin provided support "no such persons available" implies that the person does not have kin, so "no such persons available" is not a useful indicator of frequency of support. Thus, responses 6 and 7 were excluded. In addition, scores of 77 and 88 were excluded from all analyses (i.e., Japan). The item was recoded so that a high score signified higher frequency of support.

***Use of non-kin as a support resource.*** The use of non-kin as a support resource was assessed using a single item (V52 EASS 2010): "During the past 12 months, did your non-kin (friends, colleagues, or neighbors) do the following things for you when you needed it? If yes, how often?" The possible responses to this question were "Very often (1)," "Often (2)," "Sometimes (3)," "Seldom (4)," "Not at all (5)," "No, do not have such needs (6)," "No such persons available (7)," "Not asked (77)," and "Don't know/Refused (88)." Because the question specifically asked if the respondents received support when they needed it, response number 6 is not a useful indicator of frequency of support. Similarly, because the question is whether non-kin provided support "no such persons available" implies that the person does not have a person described as such, so "no such persons available" is not a useful indicator of frequency of support. Thus, responses 6 and 7 were excluded. In addition, the score of 77 was excluded from all analyses (i.e., Japan). A score of 77 was excluded from all analyses (i.e., Japan). The item was recoded so that a high score signified higher frequency of support.

*Use of mental health professionals.* The use of mental health professionals as a support resource was assessed using a single item (v55): "During the past 12 months, did your professional workers (e.g., social workers, caretakers, or therapists) do the following things for you when you needed it? If yes, how often?" The possible responses to this question were "Very often (1)," "Often (2)," "Sometimes (3)," "Seldom (4)," "Not at all (5)," "No, do not have such needs (6)," "No such persons available (7)," "Not asked (77)," and "Don't know/Refused (88)." Similar to the above two measures, responses 6 and 7 were excluded. In addition, the score of 77 was excluded from all analyses (i.e., Japan). A score of 77 was excluded from all analyses (i.e., Japan). The item was recoded so that a high score signified higher frequency of support. The item was recoded so that a high score signified higher frequency of support.

*Use of alternative medicine as a support resource.* The use of alternative medicine was assessed using three items (V46, V47, and V48 EASS 2010): "Have you ever received the following treatment during the last 12 months?" There were three categories for this question: (a) "Acupuncture or moxibustion (cupping) (v46)," "(b) Oriental herbal medicine (v47)," and (c) "Acupressure or clinical massage (48)." The possible responses to each item were "Yes (1)" and "No (2)." "Yes" answers on items V46, V47, and V48 were recoded as "1." "No" answers were recoded as "0." For analysis purposes, the sum of V46, V47, and V48 was used. Note that cupping is not included in JGSS due to its lack of popularity in Japan.

*Country.* The country/region of respondents, China, Japan, South Korea, and Taiwan was respectively coded as "1", "2", "3", and "4".

**Data Analysis**. Data screening and cleaning, including detection of anomalies, outliers, and missing value patterns, was conducted. Listwise deletion was used to treat missing cases or observations. For research questions 1, 3, 5, and 7 (rates), the following descriptive statistics were calculated: frequency, percentage, mean, median, mode, standard deviation, skewness, and kurtosis. Research Questions 2, 4, and 8 (ordinal variables) involved in the study could be treated as interval variables given their specific Likert structure and the large sample size (Jamieson, 2004). As such, differences between countries were analyzed by means of a one-way ANOVA. If the one-way ANOVA was significant, pairwise post hoc analyses were done. The Chi-square test was used to address Research Question 6, where the dependent variable is alternative medicine, and the independent variable is country.

## Results

The data for the current study comes from the EASS survey which included 10,137 respondents across the four East Asian countries: China, Japan, South Korea, and Taiwan. The Taiwanese sample (i.e., 2,199 participants) was not asked questions regarding internet/game addiction. Of the remaining sample (i.e., 7,938), 351 participants reported being perceived by others as being addicted to the internet. Thus, the analytic sample of this study included 351 respondents (118 Chinese, 116 Japanese, and 117 South Korean). Prior to conducting the primary analysis, the data were screened for missing values and outliers on the study variables as well as violations of specific statistical assumptions. Analysis of missing value patterns indicated that missing values were well below 1.8% for alternative medicine types across the three countries. The Japanese participants were not asked the questions about social support sources. Analysis of missing value patterns for support sources among Chinese and South Korean participants indicated 7.7% had a missing value for non-kin support, 9.8% for kin support and 47.7% for professional mental health services. Examination for outliers revealed that four participants had scores below 3 standard deviations from the mean for non-kin support and five participants had a score above 3 standard deviations for mental health professional service support. Given, the small number of outliers, analyses were done without having to exclude them. Although one-way ANOVA is robust to assumptions of normality, in the current study, the author tested for this assumption using *skewness* and *kurtosis* values for each of the continuous variables. Results of the analysis showed that in absolute value terms, the skewness values ranged from .04 (SE = .22) for non-kin support to 1.75 (SE = .22) for professional mental health service support. Similarly, in absolute value terms, the kurtosis values ranged from .48 (SE = .44) for non-kin support to 2.50 (SE = .44) for professional mental health service support. These ranges of values for the two statistics indicate that the continuous variables appear to range from normally distributed to relatively skewed and kurtotic.

The first research question sought to determine the usage rate of kin as a support resource among East Asians reporting internet addiction. To address this research question, we calculated descriptive statistics of kin-support as reported by the respondents. Table 1 presents how often the respondents' kin extended emotional support to the respondents. The majority (76%) of East Asian reporting internet addiction received emotional support from their kin at least sometimes in the past 12 months. More importantly, about 44% of the participants reported having received emotional support from their kin often or very often.

The second research question was concerned with whether there are statistically significant differences in the use of kin as a support resource by country. A one-way ANOVA was conducted to compare if the mean frequency of reported kin emotional support differed by country. The results indicated that there was a statistically significant difference between countries in the use of kin as a support resource, $F(1, 211) = 15.65$, $p = .00$). The Chinese respondents reported higher use of kin support ($M = 3.55$, $SD = 1.02$) than the South Korean respondents ($M = 2.93$, SD = 1.26).

Similar to the first research question, the third research question sought to determine the prevalence of non-kin as a support resource among East Asians reporting internet addiction. Again, we calculated descriptive statistics of non-kin support as indicated by the respondents. Table 1 presents how often respondents received emotional support from non-kin. The majority (81%) of East Asians reporting internet addiction received emotional support from non-kin at least sometimes in the past 12 months.

About 41% said having received emotional support from non-kin often or very often.

The fourth research was aimed at determining whether there are statistically significant differences in the use of non-kin as a support resource by country. A one-way ANOVA was conducted to compare the mean frequency of non-kin support by country. The results indicated that there were no statistically significant differences between countries in the use of non-kin as a support resource, $F(1, 216) = .02, p = .96)$.

The fifth research question sought to determine the usage rate of alternative medicine as a support resource by East Asians reporting internet addiction. We calculated percentages of respondents based on their use of three types of alternative medicine: Acupuncture/Moxibustion, Oriental herbal medicine and Acupressure/Clinical massage. Table 2 presents the number and percentage of East Asians reporting internet addiction who used alternative medicine in the past 12 months. The table shows that about one-fifth (19.5%) of East Asian reporting internet addiction used acupressure in the past 12 months. While about 14% used acupuncture, 19% used herbal medicine.

In addition to determining the usage rate of alternative medicine as a support resource by East Asians reporting internet addiction, in the sixth research question, we sought to determine whether there were statistically significant differences in the use of alternative medicine as a support resource by country. To address this research question three chi-square tests of independence were calculated: one for each alternative medicine type. Table 3 presents cross-tabulations and chi-square tests for alternative medicine by country. As can be seen from the table, there are significant differences in the use of acupuncture among the three countries, $X^2(2, 3) = 18.99$, $p=.00$. Acupuncture seems to be the most common type of alternative medicine in South Korea, with 23% of the South Korean participants reporting its use in the past 12 months. Table 3 also shows that there are small but significant differences in the use of herbal medicine among the three countries, $X^2(2, 3) = 6.29$ $p=.04$. Herbal medicine seems to be the most common type of alternative medicine in China, with 25% of the Chinese participants reporting its use in the past 12 months. Finally, there were no statistically significant differences in the use of acupressure among the three countries, $X^2(2, 3) = 1.44$, $p =.49$.

The seventh research question sought to determine the usage rate of professional mental health services as a support resource. To address this research question, we calculated descriptive statistics of the use of professional mental health services as reported by the respondents. Table 1 presents how often the respondents' received professional mental health services. The majority (68%) of East Asian reporting internet addiction did not receive professional mental health services during the last 12 months. Indeed, only about 16% reported having used professional mental health services as a support resource at least sometimes.

The last research question was concerned with whether there were statistically significant differences in the use of professional mental health services as a support resource by country. A one-way ANOVA was conducted to compare if mean frequency of reported use of professional mental health services differed by country. The results indicated that there were statistically significant differences between countries in the use of professional mental health services as a support resource, $F(1, 122) = 8.03$, $p = .01$). The Chinese respondents reported higher use of professional mental health services ($M = 1.81$, $SD = .97$) than South Korean respondents ($M = 1.31$, $SD = .93$).

## Discussion

The aim of this study was twofold. The first aim was to determine the prevalence estimates of support sources of mental healthcare assistance among East Asians reporting internet addiction. The second aim was to examine whether there are differences between countries in rates of usage of these support sources. Towards this end, this study addressed the following research questions: 1) What is the usage rate of kin as a support resource? 2) Are there statistically significant differences in the use of kin as a support resource by country? 3) What is the usage rate of non-kin as a support resource? 4) Are there statistically significant differences in the use of non-kin as a support resource by country? 5) What is the usage rate of alternative medicine as a support resource? 6) Are there statistically significant differences in the use of alternative medicine as a support resource by country? 7) What is the usage rate of professional mental health services as a support resource? 8) Are there statistically significant differences in the use of professional mental health services as a support resource by country? In this section, the results are presented and discussed in relation to previous work. Towards the end of the section, limitations of this study, and the implications of the findings for researchers and practitioners are presented.

The first research question was concerned with determining the usage rate of kin as a support resource among those reporting internet addiction. The results indicated that, among East Asians indicting internet addiction, about 76% reported having received emotional support for their mental health problems. A more intriguing finding was that about 44% of the participants reported having received emotional support from their kin *often or very often*. One plausible explanation is the scarcity of professional assistance. The ratio of professional counselors to patients is about 2.4 per 1 million in China (Chinese Psychological Society, 2004), which is a much lower ratio compared to the United States where there are 3,000 counselors per 1 million people (Hohensil, Amundson, & Niles, 2015). With little help available from professional counselors, individuals with low mental health have no option but to rely on the more accessible familial assistance. Another plausible explanation for such findings can be found in the East Asian culture. Several studies have shown that beliefs and traditional practices influence Asians' needs for social support. For instance, the Chinese core cultural values endorse family as the core unit of daily life and resource for support, harmonious social and interpersonal relations, and avoidance of extreme emotional reaction (Tseng, Lin, & Ye, 1995). Having a mental illness such as internet addiction creates a conflict with the core cultural values. As a result, mental illness is judged to be an abnormal aberrant, or a deviance (Haslam, 2005) which brings a tremendous amount of shame to the individual and to his/her family (Fabrega, 1991; Hanzawa et al., 2009; Jenni, 1999; Mellor et al., 2012). In order to avoid bringing shame to the family, individuals with internet addiction would tend to seek familial support. While both explanations contributed to the findings, the latter explanation is the most influential one. Studies have shown that the deep-rooted Confucian philosophy views mental illness as disharmony to the collective society in which it also brings shame to the family. As a result, individuals with mental illness such as addiction would prefer to preserve the family's dignity and only seek assistance within the family.

The second question was concerned with whether there are significant differences in the use of kin as a support resource by country. The findings indicated that there was a significant difference between the two East Asian countries of China and South Korea. The Chinese respondents reported higher use of kin emotional support than South Korean respondents. This result may be explained by the fact that Chinese society is less receptive to assistance from outside the family compared to their South Korean counterparts (citation needed here). One study suggested that even Korean families do not prefer someone outside of the family to care for the family member who has schizophrenia, they are likely to receive non-familial assistance (Hanzawa, 2012). In contrary, approximately 70% of Chinese individuals with schizophrenia depend almost exclusively on their families for care (Chan, 2011; Chan & Yu, 2004; Sethabouppha & Kane, 2005). Another plausible explanation is the higher availability of non-familial assistance in South Korea compared to China. It is estimated that about 19.7% of the South Korea population is self-identified as Protestant (Hong, 1999) compared to 3% in China (Fiedler, 2010). South Korean Protestant churches promote religious participation which also provides spiritual support and assistance (Lee, 2009). The latter explanation seems to be more plausible since it affirmed previous studies on religious social support and church attendance. It is suggested that religious social support can be effective in reducing mental health issues (Nooney & Woodrum, 2002; Schwadel & Falci, 2012). With the wide availability of religious support in South Korea, the need for kin support would be less than its counterpart in China.

Concerning research question #3, which focused on the usage rate of non-kin support, the results revealed that a vast majority (81%) of East Asian reporting internet addiction received emotional support from non-kin at least sometimes in the past 12 months. Previous studies evaluating the prevalence of social support among mental health patients showed that East Asians prefer seeking help from informal sources such as family and friends (Na, Ryder, & Kirmayer, 2016; Picco et al., 2016). A possible explanation for the high prevalence rate of non-kin support among the current sample could be that non-kin assistance is highly accessible. Another plausible explanation for such a finding is that it is more socially acceptable for one to seek help within one's social circle than assistance outside of the network. Studies have shown that it is a cultural norm for East Asians to expect help within one's social network (Tata & Leong, 1994). This cultural explanation is the most plausible explanation because it is in line with previous findings.

Regarding research question four, the results indicated that there was no statistically significant difference between countries in the use of non-kin as a support resource. This seems to suggest that non-kin support functions similarly across the two countries. This is the first study that compares the two countries' help-seeking behavior from non-kin among individuals reporting internet addiction.

The focus of the fifth research question was determining the usage rate of alternative medicine as a support resource by East Asians reporting internet addiction. The findings indicated that about one-fifth of the respondents reported using acupressure in the past 12 months. A similar percentage (19%) reported having used herbal medicine. About 14% reported having used acupuncture. Although the use of alternative medicine is not high compared to kin and non-kin support, the prevalence of alternative medicine in this particular study is much higher than those reported in the Western mental health literature. Unützer et al. (2000) found that about 14.5% of the U.S. respondents have reported the use of alternative medicine in the past 12 months. Among the 14.5% who have reported alternative medicine use, only 15% reported using alternative medicine to treat mental health issues (Unützer et al., 2000). The percentage used to treat internet addiction is expected to be much lower. One plausible explanation for a higher rate is the long history of alternative medicine usage in East Asian countries, and the trust of alternative medicine is much higher than in the West. East Asians have been practicing herbal medicine, acupuncture, and acupressure for more than 2,000 years (Hakverdioglu & Turk, 2006; Leslie, 1976). Another plausible explanation is that alternative medicine is highly accessible in East Asian countries. For instance, Chinese biomedicine doctors are granted the right to practice traditional medicine (Xu & Yang, 2009). Moreover, both China and South Korea have institutionalized alternative medicines into the healthcare system which increased their accessibility (Shim, 2016). Both explanations, higher credibility and accessibility, served an important role in the findings.

The sixth research question of this study addressed the issue of country-level differences in the use of alternative medicine. The findings revealed that there are some significant differences in the use of alternative medicine. In particular, a significantly higher percentage of South Korean participants reported the use of acupuncture compared to Chinese or Japanese participants. On the other hand, a significantly higher percentage of Chinese participants reported the use of herbal medicine compared to South Korean or Japanese participants. Contrary to our expectation, there are no statistically significant differences in the use of acupressure among the three countries. No prior studies have been conducted comparing East Asian countries' alternative medicine usage rates on the individual's reported internet addiction.

One plausible explanation for herbal medicine as the higher usage rate in China is the higher level of accessibility. As mentioned above, the Chinese healthcare system allowed biomedical doctors to practice traditional Eastern alternative medicine (Shim, 2016). In contrary, the Korean healthcare system separated the biomedical and traditional medicines with separate licensing systems (Shim, 2016). As a result, the chances of Chinese patients receiving traditional alternative medical treatment such as herbal medicine is higher. Another plausible explanation is affordability. Most herbal medicine in South Korea is not covered by government insurance, which reduces its affordability and accessibility to patients (Kimet al., 2012). Although the South Korean government insurance does not cover most herbal medicine treatment, it does fully cover practices such as dry cupping, moxibustion, and acupuncture (Kim et al., 2012) while Chinese patients need to pay an average of 19.32 CNY (2.8 USD) for such treatment (Zhao, 2011). Acupuncture treatment usually requires multiple sessions a month, and an average (monthly) salary of a Chinese blue-collar worker is between 5,000 and 6,214 CNY (725 USD and 901 USD) (Jing, 2015). The cost of acupuncture might not be a significant burden to most patients in China, although more research is needed on the topic. . On the other hand, Japan has the lowest usage rate of herbal medicine and acupuncture among the three countries. One plausible explanation is the lack of general recognition of traditional medicine. Unlike China and South Korea, the Japanese healthcare system does not integrate traditional medicine as the primary treatment modality. There are only vocational schools for Eastern medicine technicians (Shim, 2016). Moreover, biomedicine doctors are discouraged from integrating traditional medicine as part of a treatment plan (Shim, 2016). As a result, Japan has the lowest usage rate of alternative medicine among the three countries. The above explanations, including institutionalization of traditional Eastern medicine into the healthcare system and government insurance coverage, play an important role on the different usage rate of each country.

    The last two questions that this study addressed focused on the use of professional mental health service support. The seventh research question sought to determine the usage rate of professional mental health services as a support resource among those reporting internet addiction. The results indicated that only about 16% of respondents with internet addiction reported having used professional mental health services as a support resource at least sometimes. Previous studies have shown similar findings on professional mental health services use among Asians (Hong Kong Government Information Centre, 2001; Park et al., 2012; Xiang, Yu, Sartorius, Ungvari, & Chiu, 2012). One plausible explanation of such finding is the scarcity of trained mental health professionals to treat internet addiction. No prior study has been conducted on the availability of internet addiction assistance in East Asia. However, with only 15,000 trained psychiatrists serving a population of 1.3 billion Chinese and the licensing system for professional counselors has yet to exist in South Korea, the professional assistance available to recently emerged internet addiction would be assumed to be scarce (Ma, 2011). Another plausible explanation is the stigma of mental illness. Addiction, like other types of mental illness, is considered shameful and brings disharmony to the family (Au, 2017; Fabregas, 1991; Hanzawa et al., 2009; Jenni, 1999; Schomerus et al., 2012). Individuals with mental illness tend to choose assistance within their families and social circle instead of seeking for professional assistance (Hanzawa et al., 2009; Lam et al., 2010). Each of these explanations played an important role in the findings.

    Finally, the last research question addressed if there were statistically significant differences in the use of professional mental health services as a support resource by country. There was no prior study comparing help-seeking behavior of individuals with internet addiction between the two countries; therefore, the findings here represent a contribution to the literature. The results indicated that there were statistically significant

differences between countries in the use of professional mental health services as a support resource, with Chinese respondents reporting higher use of such services than South Koreans.  A possible reason that Chinese participants reported higher reliance on professional mental health services is that there is higher availability of professional assistance for internet addiction in China.  Another plausible explanation is that the accessibility to non-kin social support in South Korea is higher which reduces the need to seek for professional help.  A previous study on South Korean women population showed that higher social support is associated with decreased self-perceived need for professional assistance (Arnault, Gang, & Woo, 2018).  However, this explanation is inconsistent with this study's finding on non-kin usage where there are no significant statistical differences found between the two countries.  This inconsistency could be explained by the fact that the reduction of perceived need to seek professional assistance does not necessarily transform into the action of seeking assistance from non-kin.  In that case, it is the availability rather than the participation of non-kin social support which influences one's professional help-seeking behavior.  While each of the explanations mentioned here is possible, further study needs to be done to determine the cause of the findings.

There are several limitations to this study.  The first limitation is that while the data shows there is an association between the usage rate of different types of assistance among individuals reported with internet addiction, it is unclear if the usage was for treating mental health reasons.  The second limitation of the study is that the data was gathered through self-report measures.  Readers need to keep in mind that mental illness including addiction is still highly stigmatized in East Asia, so the self-report data might be subject to bias and underreport.  The third limitation of this study involves the lack of data from Taiwan and Japan which limits the representation of East Asian populations.  Finally, the scope of this study was limited to understanding the prevalence and country-level difference in support sources; thus, the study did not determine how such support sources can buffer the effect of internet addiction.

Despite these limitations, findings of the current study have important implications for both conceptualization of treatment and treatment practice.  The scientific study of internet addiction is in its infancy.  This study offered a culturally sensitive perspective on help-seeking resource preferences for internet addiction in East Asia. First, these findings can be used for the development of a more holistic approach to treating internet addiction. Second, the high rates of assistance from kin, non-kin, and alternative medicine warrant a closer look for a more holistic approach to East Asian patients with internet addiction issues. Family members, friends, and alternative medicines should be part of the intervention process. Third, since kin and non-kin play such an important role in the recovery process, systematic psychoeducation should be developed to improve treatment outcome. Finally, East Asian healthcare systems should educate the public about internet addiction to reduce stigma. Once stigma is reduced, help-seeking behavior might be increased across the spectrum.

Table 1

*Frequency and Percentage of Support Sources*

| Rating | Kin Support Frequency | % | Non-Kin Support Frequency | % | Professional Support Frequency | % |
|---|---|---|---|---|---|---|
| Not at all | 22 | 10.4 | 14 | 6.5 | 84 | 68.3 |
| Seldom | 29 | 13.7 | 28 | 12.9 | 19 | 15.4 |
| Sometimes | 67 | 31.6 | 86 | 39.6 | 13 | 10.6 |
| Often | 61 | 28.8 | 65 | 30.0 | 4 | 3.3 |
| Very Often | 33 | 15.6 | 24 | 11.1 | 3 | 2.4 |
| Total | 212 | 100 | 217 | 100 | 123 | 100.0 |

Table 2

*Frequency and Percentage of Respondents Using Alternative Medicine*

| Response | Acupuncture | | Herbal medicine | | Acupressure | |
|---|---|---|---|---|---|---|
| | Frequency | % | Frequency | % | Frequency | % |
| No | 299 | 86.2 | 281 | 80.7 | 280 | 80.5 |
| Yes | 48 | 13.8 | 67 | 19.3 | 68 | 19.5 |
| Total | 347 | 100 | 348 | 100 | 348 | 100 |

Table 3

*Cross-Tabulations and Chi-square Test Results for Alternative Medicine by Country*

| Alternative Medicine Type | China | Japan | South Korea | Chi-Square Tests of Independence |
|---|---|---|---|---|
| Acupuncture | | | | $X^2(2, 3) = 18.994$, $p = .000$ |
| No  $n$ (%) | 97 (85.1%) | 112 (96.6%) | 90 (76.9%) | |
| Yes  $n$ (%) | 17 (14.9%) | 4 (3.4%) | 27 (23.1%) | |
| Herbal medicine | | | | $X^2(2, 3) = 6.291$, $p = .043$ |
| No  $n$ (%) | 87 (75.0%) | 101 (87.8%) | 93 (79.5%) | |
| Yes  $n$ (%) | 29 (25.0%) | 14 (12.2%) | 24 (20.5%) | |
| Acupressure | | | | $X^2(2, 3) = 1.436$ $p = .488$ |
| No  $n$ (%) | 90 (77.6%) | 92 (80.0%) | 98 (83.8%) | |
| Yes  $n$ (%) | 26 (22.4%) | 23 (20.0%) | 19 (16.2%) | |